\documentclass[twocolumn]{autart}    

\usepackage{graphicx}          
\usepackage{amsfonts}
\usepackage{amssymb}
\usepackage[fleqn]{amsmath}
\usepackage{tikz}
\usepackage{verbatim}
\usetikzlibrary{arrows.meta, positioning, quotes, calc, shapes.geometric}

\tikzset{
	block/.style = {draw, rectangle,
		minimum height=1cm,
		minimum width=2cm},
	input/.style = {coordinate,node distance=1cm},
	output/.style = {coordinate,node distance=4cm},
	arrow/.style={draw, -latex,node distance=2cm},
	pinstyle/.style = {pin edge={latex-, black,node distance=2cm}},
	sum/.style = {draw, circle, node distance=1cm},
}

\newcommand{\R}{\mathbb{R}}
\newcommand{\Z}{\mathbb{Z}}

\newcommand{\relu}{\operatorname{ReLU}}

\newtheorem{asm}{Assumption}

\definecolor{agentcolor1}{rgb}{0, 0.4470, 0.7410}
\definecolor{agentcolor2}{rgb}{0.8500, 0.3250, 0.0980}
\definecolor{agentcolor3}{rgb}{0.9290, 0.6940, 0.1250}
\definecolor{agentcolor4}{rgb}{0.4940, 0.1840, 0.5560}
\definecolor{agentcolor5}{rgb}{0.4660, 0.6740, 0.1880}
\definecolor{dhs}{RGB}{0,100,255}

\definecolor{hyj}{RGB}{0,50,250}

\definecolor{red}{RGB}{255,10,10}

\begin{document}

\begin{frontmatter}

\title{A Note on Emergent Behavior in Multi-agent Systems Enabled by Neuro-spike Communication \thanksref{footnoteinfo}} 
\thanks[footnoteinfo]{
Corresponding author H. Shim.
}

\author[ASRI]{Hyeonyeong Jang}\ead{hyjang@cdsl.kr},    
\author[ASRI]{Donghyeon Song}\ead{dhsong@cdsl.kr},    
\author[ASRI]{Jin Gyu Lee}\ead{jingyu.lee@snu.ac.kr},  
\author[ASRI]{Hyungbo Shim}\ead{hshim@snu.ac.kr}       

\address[ASRI]{ASRI, Department of Electrical and Computer Engineering, Seoul National University, Seoul, Republic of Korea}  
          

\begin{abstract}                          
In this note, we show emergent behavior can arise even when the agents interact solely through spikes.
Rather than exchanging real-valued states, each agent encodes its own state by bio-inspired neuromorphic amplifiers and transmits signed spikes, so that each firing carries a single bit and the traffic is intermittent and asynchronous.
We model the resulting network as a hybrid system and introduce a coordinate that is invariant under the spike-induced jumps, which reduces the convergence analysis to a flow dynamics without jumps.
On this coordinate, we prove that the agents practically synchronize and collectively follow the blended dynamics that characterizes an emergent behavior, and we derive explicit ultimate error bounds for synchronization and for the expected behavior.
The bound reveals a certified accuracy--threshold trade-off and yields closed-form tuning rules under a minimum-threshold or prescribed-accuracy constraints.
A distributed median solver is tested using neuro-spike communication.
\end{abstract}

\end{frontmatter}

\section{Introduction}

Emergent behavior---the phenomenon in which local interactions give rise to a collective behavior not exhibited by any individual agent---is a central theme in neuroscience and complex systems theory.
Biological systems often achieve such coordination through discrete pulse-like events: fireflies synchronize through brief flashes~\cite{mirollo1990synchronization}, and neural systems process information through all-or-none pulses, called ``spikes'', which encode
information primarily in their timing rather than in their amplitude~\cite{rieke1999spikes}.

In engineering, by contrast, many conventional synchronization protocols for multi-agent systems assume continuous access to the real-valued states of neighboring agents.
Event-triggered strategies alleviate this requirement by communicating only at selected triggering instants~\cite{ding2017overview,nowzari2019event}, but the transmitted messages often remain real-valued packets; hence communication becomes intermittent while the payload of each transmission can remain substantial, which is particularly restrictive in severely bandwidth-constrained settings such as underwater acoustic networks~\cite{akyildiz2005underwater}.
This motivates a central question: \emph{can heterogeneous multi-agent systems exhibit emergent behavior when their inter-agent communication is restricted entirely to spikes, and if so, what collective behavior emerges, through what mechanism, and under what conditions?}

To address this question, we build on the blended dynamics theory~\cite{kim2015robustness,lee2020tool}, which characterizes the emergent behavior of strongly coupled heterogeneous networks.
Consider a network of $N$ agents described by
\begin{equation}\label{eq: Heterogeneous MAS}
\dot x_i = f_i(t,x_i) + u_i,\quad i\in\mathcal{N}:=\{1,\dots,N\},
\end{equation}
where $x_i\in\mathbb{R}^n$ and $u_i\in\mathbb{R}^n$ are the state and the input of agent~$i$, respectively.
Suppose that these agents are diffusively coupled through $u_i = k\sum_{p\in\mathcal{N}_i}(x_p - x_i)$, where $\mathcal{N}_i \subseteq \mathcal{N}$ is the set of agents sending their states to agent~$i$.
Provided that the \emph{blended dynamics}
\begin{equation}\label{eq: Blended dynamics}
\dot s = \frac{1}{N}\sum_{i\in\mathcal{N}}f_i(t,s) =: \bar f(t,s)
\end{equation}
is stable in an appropriate sense, the theory guarantees that, with a sufficiently large gain $k$, the heterogeneous agents practically synchronize and collectively follow the blended dynamics~\eqref{eq: Blended dynamics}.
The resulting collective trajectory is thus generated by an averaged vector field $\bar f$ that need not coincide with any individual vector field $f_i$.
What this guarantee presumes, however, is again the continuous exchange of real-valued states that we claim to remove.

To replace this continuous communication channel, we employ a bio-inspired device introduced and analyzed in~\cite{petri2024analysis,petri2025spiking}, which we call the \emph{neuromorphic amplifier} (NM-AMP).
Composed of a pair of integrate-and-fire neurons, the NM-AMP separately accumulates the positive and negative parts of a scalar input and emits a signed spike of fixed amplitude $\alpha$ whenever the corresponding membrane potential reaches a threshold $\Delta$.
Each firing therefore carries only a one-bit sign, while its timing reflects the accumulated input.
Despite carrying no magnitude information, the NM-AMP emulates the continuous gain $k := \alpha/\Delta$, in the sense that the integral of the amplification error remains uniformly bounded.
Existing results exploit this property to stabilize single linear time-invariant systems; moving beyond this, we establish that collective behavior can emerge in heterogeneous networks interacting solely through spikes.

We now deploy the NM-AMP to realize the diffusive coupling through spikes.
A direct emulation, obtained by feeding $\sum_{p\in\mathcal{N}_i}(x_p-x_i)$ into an NM-AMP, would undermine the purpose, since forming the input already requires continuous access to the real-valued states of the neighbors.
Instead, we exploit the decomposition 
\begin{equation}\label{eq: Decoupled diffusive coupling}
u_i = -k|\mathcal{N}_i|x_i + k\sum_{p\in\mathcal{N}_i}x_p.
\end{equation}
The self-feedback term $-k|\mathcal{N}_i|x_i$ can be computed locally using only the state and the degree of agent~$i$.
To emulate the neighbor-dependent term, each agent feeds its own state components into independent NM-AMPs and transmits only the resulting signed spikes, which the receiving agents apply directly to their own states.
Thus, the neighbor-dependent part of the coupling is realized entirely through spikes, whereas the continuous self-feedback requires no network communication.
The remaining question is whether this spike-based implementation preserves practical synchronization and the emergent behavior predicted as the blended dynamics.

To answer this question, we formulate the resulting closed-loop network as a hybrid system and establish its well-posedness.
We then introduce a jump-invariant coordinate that removes jump contributions from the convergence estimate, reducing the convergence analysis to the flow dynamics.
Under suitable connectivity, regularity, and contraction conditions, we prove that every maximal solution is complete and Zeno-free and that all agents remain ultimately close to a common trajectory generated by the blended dynamics, with an explicit error bound.
The explicit bound reveals how the spike amplitude and threshold affect the guaranteed synchronization accuracy and yields closed-form tuning rules under minimum-threshold or prescribed-accuracy constraints.

We illustrate the framework on a distributed median solver~\cite{lee2020fully}, in which each agent knows only its own reference value and the median emerges as the collective behavior.
The example is deliberately outside the scope of the convergence theorem: its blended dynamics is not contractive, yet the spike-coupled network still exhibits the predicted emergent behavior, in agreement with the extensions of blended dynamics theory to broader stability notions~\cite{lee2020tool,lee2020fully}.

The paper is organized as follows.
Section 2 collects preliminaries on the communication topology, hybrid dynamical systems, and the neuromorphic amplifier.
Section 3 presents the spike-based coupling and its hybrid realization, the jump-invariant coordinate, the convergence analysis, and the resulting accuracy--communication trade-off.
Section 4 reports the numerical example, and Section 5 concludes.

\textbf{Notation:} 
Let $\mathbb{R}$ and $\mathbb{Z}$ denote the sets of real numbers and integers, respectively. For a set $E \subseteq \mathbb{R}$ and a constant $c \in \mathbb{R}$, we define $E_{\ge c} := \{x \in E \mid x \ge c\}$.
The rectified linear unit is defined by $\relu(x) := \max(0, x)$, where the maximum is applied componentwise for vector-valued arguments.
For matrices $M_1, \dots, M_n$, possibly of different dimensions, $\operatorname{diag}(M_1, \dots, M_n)$ denotes the block diagonal matrix with diagonal blocks $M_1, \dots, M_n$.
For vectors $v_1, \dots, v_n$, the vertical concatenation of $v_i$'s is denoted by $[v_1;\cdots;v_n]$.
For example, the vector of all ones is defined by $\mathbf{1}_N := [1;\cdots;1]\in\mathbb{R}^N$.
For a vector $v$ and a matrix $M$, $\|v\|$ and $\|M\|$ denote the Euclidean norm and the induced matrix 2-norm, respectively.
The vector $1$-norm and $\infty$-norm are denoted by $\|\cdot\|_1$ and $\|\cdot\|_\infty$, respectively.
The partial derivative operator of the $i$-th argument is denoted by $\partial_i$.
For square matrices of identical dimension $M_1$ and $M_2$, $M_1\prec M_2$ ($M_1\preceq M_2$) means $M_2 - M_1$ is symmetric positive definite (semidefinite).
For a vector $v$ and a matrix $Q\succ0$, we define $\|v\|_Q = \sqrt{v^\top Q v}$.
The Kronecker product is denoted by $\otimes$.
For an arbitrary matrix $M$, $M \otimes I_n$ is denoted by $M_{\otimes n}$.
We denote the upper Dini derivative by $D^+$.

\section{Preliminaries}

\subsection{Communication topology}

The communication topology of the network governed by \eqref{eq: Heterogeneous MAS} is represented by a directed graph $\mathcal{G} = (\mathcal{N},\mathcal{E})$,
where $\mathcal{N} = \{ 1,2,\dots,N \}$ is the set of agents, and $\mathcal{E}\subseteq\mathcal{N}\times\mathcal{N}$ is the edge set.
The pair $(p,i)$ belongs to $\mathcal{E}$ if and only if $p \in \mathcal{N}_i$.
The \emph{adjacency matrix} $\mathcal{A}=[a_{ip}]\in\mathbb{R}^{N\times N}$ of $\mathcal{G}$
is defined such that $a_{ip}=1$ if $(p,i) \in \mathcal{E}$,
and $a_{ip}=0$ otherwise.
Accordingly, the $i$-th row of $\mathcal{A}$ identifies the agents that send information to agent~$i$, whereas the $i$-th column identifies the agents that receive information from agent~$i$.
The \emph{degree matrix} $\mathcal{D}=[d_{ip}]\in\mathbb{R}^{N\times N}$ of $\mathcal{G}$
is a diagonal matrix defined such that $d_{ii} = \sum_{p\in\mathcal{N}}a_{ip}$,
and $d_{ip} = 0$ for $i\not=p$.
The \emph{Laplacian matrix} $\mathcal{L}\in\mathbb{R}^{N \times N}$ of $\mathcal{G}$
is defined as $\mathcal{L} = \mathcal{D} - \mathcal{A}$. Note that by the definition, $\mathcal{L}\mathbf 1_N=0$.

\begin{asm}\label{asm: connected}
The graph $\mathcal G=(\mathcal N,\mathcal E)$ has no self-loop, is strongly connected,
and is balanced, i.e., $\mathbf 1_N^\top \mathcal{L}=0$.
\end{asm}
Let $\mathcal{L}_s:=(\mathcal{L}+\mathcal{L}^\top)/2$, which is the Laplacian of the mirror graph of
$\mathcal G$ and is therefore symmetric positive semidefinite with a simple
zero eigenvalue~\cite{olfati2007consensus}; we denote its smallest positive eigenvalue by $\lambda_2$.

\subsection{Hybrid systems}

We consider the hybrid system $\mathcal{H} = (C, F,D, G)$ in the framework of \cite{goebel2012hybrid}
\begin{equation*} \label{eq:hybrid_basic}
\mathcal{H}:\begin{cases}\begin{aligned}
\dot \chi &\in F(\chi)  &\chi &\in C,\\
\chi^+ &\in G(\chi)  &\chi &\in D,
\end{aligned}\end{cases}
\end{equation*}
with $\chi \in \R^\rho$ being the state, where $C\subseteq \R^{\rho}$ is the flow set, $D \subseteq\R^\rho$ is the jump set, $F:\R^\rho \rightrightarrows \R^\rho$ is the flow map, and $G:\R^\rho \rightrightarrows \R^\rho$ is the jump map.
A solution to $\mathcal{H}$ is defined over a subset of $\R_{\ge0} \times \Z_{\ge0}$, called hybrid time domain~\cite[Chapter~2]{goebel2012hybrid}. 
A solution is maximal if it is not a proper truncation of another solution, complete if its hybrid time domain is unbounded, and Zeno if it has infinitely many jumps within a finite amount of flow time.

\subsection{Neuromorphic amplifier}

We now describe a pair of integrate-and-fire neurons, which we call the \emph{neuromorphic amplifier}~(NM-AMP), depicted in Fig.~\ref{fig: NM-AMP}, and introduced in~\cite{petri2024analysis}.

For a scalar input $u\in\mathbb{R}$, the internal neuronal dynamics of the NM-AMP is described by
\begin{equation}\label{eq: neuro-amp}
\begin{cases}
\dot{\xi}_{\ell}=\relu(\ell \cdot u),
    & \xi_{\ell}\in[0,\Delta],\\
\xi_{\ell}^{+}=\xi_{\ell}-\Delta,
    & \xi_{\ell}=\Delta,
\end{cases}
\qquad \ell\in\{+1,-1\},
\end{equation}
where $\xi_{\ell}\in\mathbb{R}$ is the membrane potential of neuron $\ell$, and $\Delta\in\mathbb{R}_{>0}$ is the firing threshold.
Throughout the paper, for any family $\{z_{\ell}\}_{\ell\in\{+1,-1\}}$, we abbreviate $z_{+1}$ and $z_{-1}$ as $z_{+}$ and $z_{-}$, respectively.
A sign in a subscript labels the neuronal polarity, whereas the superscript $+$ denotes the post-jump value.

The two neurons process opposite polarities of the input: the positive neuron integrates $\relu(u)$, whereas the negative neuron integrates $\relu(-u)$.
When $\xi_{\ell}=\Delta$, a firing jump is enabled, and the reset after the firing in \eqref{eq: neuro-amp} gives $\xi_{\ell}^{+}=0$.
Let $\mathcal{J}_{\ell}\subset \mathbb{Z}_{\ge0}$ denote the set of firing indices of neuron $\ell$, and let $t_{\ell,j}$, for $j\in\mathcal{J}_\ell$, denote its $j$-th firing time.
For a spike amplitude $\alpha\in\mathbb{R}_{>0}$, the output of the NM-AMP is the signed impulse train
\begin{equation}\label{eq: total output of NM-AMP}
y(t) = \sum_{\ell\in\{+1,-1\}} \sum_{j\in\mathcal{J}_{\ell}} \ell\cdot\alpha\delta(t-t_{\ell,j}),
\end{equation}
where $\delta$ denotes the Dirac delta distribution.
Accordingly, \eqref{eq: total output of NM-AMP} is understood in the distributional sense rather than as an ordinary function.

It was shown in \cite{petri2025spiking} that the output $y$ approximates $ku$ with $k:=\alpha/\Delta$, in the
sense that the amplification error $y-ku$ has a uniformly bounded integral;
accordingly, we call $k$ the gain of the NM-AMP \eqref{eq: neuro-amp}--\eqref{eq: total output of NM-AMP}.
In signal-processing terms, the NM-AMP is a bipolar, clock-free first-order $\Sigma\Delta$ encoder~\cite{yoon2016lif}.

\begin{figure}[t] 
\centering
\begin{tikzpicture}[
>=Latex, 
thick,   
node distance=1.0cm and 1.5cm, 
block/.style = {draw, thick, minimum width=1.6cm,
minimum height=0.9cm, align=center},
gain/.style = {draw, thick, isosceles triangle,
isosceles triangle apex angle=60, 
shape border rotate=0, 
minimum height=1cm,
inner sep=1pt, align=center},
sum/.style = {draw, circle, inner sep=1pt, minimum size=0.6cm},
scale=0.6, transform shape
]
\coordinate (u_start) at (0,0);
\node[xshift=0.75cm,above=0.125cm] at (u_start) {\large$u$};
\coordinate (split) at (1.5,0);
\node[block, right=of split, xshift=2.5cm, yshift=1.25cm] (lif1) {Neuron $+$ \\ $(\xi_+, \alpha, \Delta)$};
\coordinate (up_end) at (11.1,1.25);
\node[xshift=-3.25cm, above=0.125cm] at (up_end) {}; 
\node[gain, right=of split, yshift=-1.25cm] (gain1) {$-1$};
\node[block, right=of gain1] (lif2) {Neuron $-$ \\ $(\xi_-, \alpha, \Delta)$};
\node[gain, right=of lif2] (gain2) {$-1$}; 
\coordinate (down_end) at (11.1,-1.25);
\node[xshift=-3.25cm, above=0.125cm] at (down_end) {}; 
\node[sum, yshift = -1.25cm] at (up_end) (sum_node) {$+$}; 
\coordinate (y_end) at ($(sum_node.east)+(1.5,0)$);
\node[xshift=-0.75cm,above=0.125cm] at (y_end) {\large$y$};
\draw[thick] (u_start) -- (split);
\draw[thick] (split) -- (split |- lif1.west); 
\draw[thick,-Latex] (split |- lif1.west) -- (lif1); 
\draw[thick] (lif1.east) -- (up_end); 
\draw[thick,-Latex] (up_end) -- (sum_node.north); 
\draw[thick] (split) -- (split |- gain1.west); 
\draw[thick,-Latex] (split |- gain1.west) -- (gain1); 
\draw[thick,-Latex] (gain1) -- (lif2);
\draw[thick,-Latex] (lif2) -- (gain2);
\draw[thick] (gain2.east) -- (down_end);
\draw[thick,-Latex] (down_end) -- (sum_node.south); 
\draw[thick,-Latex] (sum_node.east) -- (y_end);
\end{tikzpicture}
\caption{Neuromorphic amplifier}
\label{fig: NM-AMP}
\end{figure}

\section{Synchronization via neuro-spike communication}

The primary objective of this paper is to establish a theoretical foundation for emergent behavior in heterogeneous multi-agent systems~\eqref{eq: Heterogeneous MAS} whose inter-agent information exchange is entirely spike-based. 
Before presenting the main synchronization protocol and its analysis, we impose a regularity condition on the individual nonlinear dynamics to ensure the well-posedness of the system and the existence of solutions.

\begin{asm}\label{asm: Regularity}
The vector field $f_i:[0,\infty)\times\mathbb{R}^n \to \mathbb{R}^n$ is continuous in $t$, continuously differentiable with respect to $x_i$ and globally $L_f$-Lipschitz\footnote{Although we restrict our attention to globally Lipschitz dynamics for simplicity, the present framework admits an extension to the locally Lipschitz case, working on a compact set, as demonstrated in~\cite{lee2020tool}.} with respect to $x_i$ uniformly in $t$, and $\|f_i(t,0)\|$ is uniformly bounded by $m_0$ for all $t\ge0$.
\end{asm}

\subsection{Spike-based coupling}

We now implement the decomposition~\eqref{eq: Decoupled diffusive coupling} using NM-AMPs.
The self-feedback term $-k|\mathcal{N}_i|x_i$ is computed locally, so only the neighbor-dependent contribution must be realized through inter-agent communication.
For each agent~$i$ with $x_i=[x_{i,1};\cdots;x_{i,n}]\in\mathbb{R}^n$,
we assign one NM-AMP to each state component.
All NM-AMPs use common parameters $\alpha$ and $\Delta$, and hence $k=\alpha/\Delta$.
For each $m\in\{1,\ldots,n\}$, the $m$-th NM-AMP of agent~$i$ is driven by $x_{i,m}$, has membrane potentials
$\xi_{\ell,i,m}$, $\ell\in\{+1,-1\}$, and generates the signed impulse train $y_{i,m}$ defined as in~\eqref{eq: total output of NM-AMP}.
Let $y_i:=[y_{i,1};\cdots;y_{i,n}]$ denote the resulting vector-valued spike train transmitted by agent~$i$.
Agent~$i$ sums the spike trains received from its in-neighbors and formally applies the input
\begin{equation}\label{eq: neuro-spike input of MAS}
u_i = -k|\mathcal{N}_i|x_i + \sum_{p\in\mathcal{N}_i}y_p.
\end{equation}

Define the stacked state, spike train, and vector field by
$x := [x_1;\cdots;x_N]\in\mathbb{R}^{nN}$, $y := [y_1;\cdots;y_N]$, and $f(t,x) := [f_1(t,x_1);\cdots;f_N(t,x_N)]$, respectively.
Then the formal impulsive representation of the coupled system is described by
\begin{equation}\label{eq: Total NM MAS}
\dot x = f(t,x) - k\mathcal{D}_{\otimes n}x + \mathcal{A}_{\otimes n}y.
\end{equation}
Note that since $y$ is a train of Dirac impulses, \eqref{eq: Total NM MAS} is not a differential equation in the classical sense;
rather, it compactly represents continuous evolution between firing instants together with the state jumps induced by the received spikes.
The hybrid system constructed in the next subsection provides a rigorous realization of this formal equation.

\begin{rem}
When $x_i$ is an algorithmic or controller state that can be updated discretely, the Dirac impulses in~\eqref{eq: Total NM MAS} need not be physically generated.
A firing of the $m$-th NM-AMP of agent~$i$ with polarity $\ell$ is implemented as the instruction
\begin{equation*}
    x_{p,m}^{+}=x_{p,m}+\ell\cdot\alpha, \quad \text{for each }p\text{ such that }i\in\mathcal{N}_p.    
\end{equation*}
If $x_i$ is instead a physical plant state that cannot be reset instantaneously, the hybrid system in the next subsection should be regarded as an ideal impulsive model; the treatment of finite-duration actuation pulses requires a separate robustness analysis.
\end{rem}

\subsection{Hybrid realization}

For each $\ell\in\{+1,-1\}$, define $\xi_{\ell,i} = [\xi_{\ell,i,1}; \cdots ; \xi_{\ell,i,n}]\in\mathbb{R}^{n}$, $\xi_\ell = [\xi_{\ell,1} ; \cdots ; \xi_{\ell,N}]\in\mathbb{R}^{nN}$, and let $\xi = [\xi_+ ; \xi_-]\in\mathbb{R}^{2nN}$.
To render the time-dependent vector fields autonomous, we introduce a timer state $\tau$ and define $q = [x; \xi ; \tau]\in\mathbb{R}^{3nN+1}$.
The formal impulsive model~\eqref{eq: Total NM MAS} admits the hybrid realization $\mathcal{H}=(C,F,D,G)$, with
\begin{equation}\label{eq: hybrid MAS}
\begin{aligned}
    &C = \mathbb{R}^{nN} \times [0,\Delta]^{2nN} \times \mathbb{R}_{\ge0}, \\
    &D= \bigl\{ q \in C \mid \ \Theta(q)\neq\varnothing \bigr\}   \\
    &F(q) =
\left\{
\begin{bmatrix}
f(\tau,x) - k\mathcal{D}_{\otimes n}x \\ \relu(x) \\ \relu(-x) \\ 1
\end{bmatrix}
\right\}, \\
&G(q) = \bigl\{ q + g_{\ell,i,m} \mid (\ell,i,m) \in \Theta(q) \bigr\}
\end{aligned}
\end{equation}
where $\Theta(q) = \{(\ell,i,m) \mid \xi_{\ell,i,m} = \Delta\}$, and
\begin{equation*}
    g_{\ell,i,m} = \bigl[\ell\cdot\alpha([\mathcal{A}]_i\otimes e_{m|n}) ; -\Delta ( b_\ell\otimes e_{i|N}\otimes e_{m|n} ) ; 0\bigr],
\end{equation*}
where $[\mathcal{A}]_i$ denotes the $i$-th column of $\mathcal{A}$, $b_+ = [1;0]$, $b_- = [0;1]$, and $e_{X|Y}\in\mathbb{R}^Y$ is the $X$-th standard basis vector of $\mathbb{R}^Y$.

The data of $\mathcal{H}$ have the following interpretation. 
Along flows, each agent evolves according to $\dot{x}_i=f_i(\tau,x_i)-k|\mathcal{N}_i|x_i$, using only its intrinsic dynamics and locally computable self-feedback.
Meanwhile, its positive and negative membrane potentials integrate the corresponding components of its own state according to $\dot{\xi}_{\ell,i} = \relu(\ell\cdot x_i)$.
Thus, no information from the neighbors is used during flows. 
The timer state evolves as $\dot{\tau}=1$. 
As a \emph{standing assumption}, we set the initial condition $\tau(0,0) = 0$ without loss of generality, hence giving $\tau(t,j) = t$. 
Note that this timer state $\tau$ is not a communicated or synchronized state; it only records the physical-time argument already present in $f_i(t,x_i)$.

The neighbors enter only through jumps, which are enabled whenever
$\Theta(q)\neq\varnothing$.
The increment $g_{\ell,i,m}$ acts on two blocks of $q$.
It decrements the firing potential, $\xi_{\ell,i,m}^+=\xi_{\ell,i,m}-\Delta$, leaving all
others unchanged.
As $[\mathcal{A}]_i$ collects the agents that receive information from agent~$i$,
the increment further sets $x_{p,m}^+=x_{p,m}+\ell\cdot\alpha$ for every $p$ with $i\in\mathcal{N}_p$,
delivering the spike $y_{i,m}$ in \eqref{eq: neuro-spike input of MAS}.
If $\Theta(q)$ contains multiple indices, then $G(q)$ is not a singleton but a multi-element set and each jump processes one active firing; the remaining firings are processed consecutively.
Consequently, solutions may be nonunique, but all results below hold for every maximal solution.

The resulting communication architecture has the following notable features.
First, communication is intermittent and asynchronous: each of the $2nN$ neurons fires according to its own threshold condition, without requiring a common transmission schedule.
This contrasts with the ideal diffusive coupling in~\eqref{eq: Decoupled diffusive coupling}, which assumes continuous access to real-valued neighbor states.
Second, the spike amplitude $\alpha$ is fixed a priori and known to all agents, so no magnitude value is transmitted; each componentwise firing carries only a one-bit polarity payload.
Thus, spikes from different neighbors are aggregated without using the sender identity in the numerical update.

\begin{prop}\label{prop: well-posedness}
    Under Assumption~\ref{asm: Regularity}, the hybrid system \(\mathcal H=(C,F,D,G)\) in~\eqref{eq: hybrid MAS} is well-posed in the sense of~\cite{goebel2012hybrid}.
\end{prop}
\begin{pf}
The flow set $C$ is closed, and so is $D$, being the intersection of $C$ with a finite union of hyperplanes $\{\xi_{\ell,i,m}=\Delta\}$.
By Assumption~\ref{asm: Regularity} and the continuity of $\relu$, $F$ is continuous and locally bounded relative to $C$, with nonempty convex singleton values.
The graph of $G$ is the finite union, over $(\ell,i,m)$, of the graphs of the affine maps $q\mapsto q+g_{\ell,i,m}$ restricted to the closed sets $D\cap\{\xi_{\ell,i,m}=\Delta\}$, and is therefore closed.
Moreover, since the increments are constant and finite in number, $G$ is also locally bounded relative to $D$, and it has nonempty values on $D$.
Its closed graph and local boundedness then imply that $G$ is outer semicontinuous relative to $D$.
The hybrid basic conditions in~\cite[Assumption~6.5]{goebel2012hybrid} therefore hold, and well-posedness follows  from~\cite[Theorem~6.30]{goebel2012hybrid}.
\hfill$\square$
\end{pf}

\subsection{Jump-invariant coordinate}

In general, a solution to $\mathcal{H}$ has a discontinuous $x$-component, so a Lyapunov analysis for $\mathcal{H}$ would have to account for both flows and jumps.
The jump condition is the delicate part: each firing displaces the receiving states by $\pm\alpha$, so a Lyapunov function expressed in $x$ may increase or decrease at a jump, while the number of firings on a finite interval is not bounded a priori; the increments can therefore be neither signed nor dismissed as finitely many.
The jumps are, however, strongly correlated: the firing that adds $\ell\alpha$ to the neighbors of agent~$i$ simultaneously decrements $\xi_{\ell,i,m}$ by $\Delta$.
Since $k=\alpha/\Delta$, a particular linear combination of $x$ and $\xi$ is unaffected by this exchange: writing $\zeta:=\xi_+-\xi_-\in\mathbb{R}^{nN}$, we set
\begin{equation}\label{eq: jump invariant coordinate}
  w := x + k\mathcal{A}_{\otimes n}\zeta .
\end{equation}

\begin{lem}\label{lem: jump invariance}
Along every solution to $\mathcal{H}$ with initial condition $q(0,0) \in C \cup D$, the coordinate \eqref{eq: jump invariant coordinate}
satisfies
\begin{enumerate}
\item[(i)] $w^+=w$ at every jump;
\item[(ii)] $\dot w = f(\tau,x)-k\mathcal{L}_{\otimes n}x$ along flows;
\item[(iii)] $\|w-x\|\le  \|\mathcal{A}\|\sqrt{nN} \alpha =: c_w \alpha$.
\end{enumerate}
\end{lem}
\begin{pf}
Let $(\ell,i,m)\in\Theta(q)$.
Since $\zeta = \xi_+-\xi_-$, the increment $g_{\ell,i,m}$ gives $\zeta^+-\zeta = -\ell\Delta\,(e_{i|N}\otimes e_{m|n})$, so that, using $\mathcal{A}_{\otimes n}(e_{i|N}\otimes e_{m|n}) = [\mathcal{A}]_i\otimes e_{m|n}$,
\begin{equation*}
  w^+-w=\ell(\alpha-k\Delta)\left([\mathcal{A}]_i\otimes e_{m|n}\right)=0,
\end{equation*}
which proves (i).
Along flows, $\dot\zeta=\relu(x)-\relu(-x)=x$, and, with $\mathcal{L}=\mathcal{D}-\mathcal{A}$,
\begin{equation*}
  \dot w=f(\tau,x)-k\mathcal{D}_{\otimes n}x+k\mathcal{A}_{\otimes n}x
        =f(\tau,x)-k\mathcal{L}_{\otimes n}x,
\end{equation*}
which is (ii).
Finally, every solution satisfies
$\xi\in[0,\Delta]^{2nN}$, hence $\|\zeta\|_\infty\le\Delta$ and
$\|\zeta\|\le\Delta\sqrt{nN}$. Since $\|\mathcal{A}_{\otimes n}\|=\|\mathcal{A}\|$, we obtain $\|w-x\|\le k\|\mathcal{A}\|\,\|\zeta\|\le  \|\mathcal{A}\|\sqrt{nN}\alpha$,
which is (iii). \hfill $\square$
\end{pf}

In this coordinate, any Lyapunov-type quantity depending only on $w$ is unchanged at jumps, by Lemma~\ref{lem: jump invariance} (i).
Lemma~\ref{lem: jump invariance} (ii) shows that the flow dynamics of $w$ coincide with those of the ideal diffusively coupled system evaluated at the physical state $x$, rather than $w$ itself; Lemma~\ref{lem: jump invariance} (iii) bounds this mismatch.

We next use Lemma~\ref{lem: jump invariance} to establish completeness and the absence of Zeno behavior for all maximal solutions.

\begin{prop}
\label{prop: complete-zeno}
Under Assumption~\ref{asm: Regularity}, for every $\alpha,\Delta>0$ and
$k=\alpha/\Delta$, every maximal solution to $\mathcal H$ in \eqref{eq: hybrid MAS} with
$q(0,0)\in C\cup D$ and $\tau(0,0)=0$ is complete and Zeno-free.
\end{prop}

\begin{pf}
Fix a maximal solution $q=(x,\xi,\tau)$ and set $\gamma_k:=L_f+k\|\mathcal L\|$.
Since $\tau(t,j)=t$, Assumption~\ref{asm: Regularity} and Lemma~\ref{lem: jump invariance} imply, along flows,
\begin{align*}
  \|\dot w\|
  &\le \sqrt N m_0+\gamma_k\|x\|\\
  &\le \gamma_k\|w\|+\sqrt N m_0+\gamma_k c_w\alpha.
\end{align*}
Since $w$ is unchanged at jumps, the comparison lemma~\cite{khalil2002nonlinear}
applied to $\|w\|$ on successive flow intervals implies that
$w$ is bounded on every bounded time interval, independently of the number of jumps.
Together with $\|x-w\|\le c_w\alpha$, this shows that,
for every $T>0$, there exists $M_T<\infty$ such that
\begin{equation}\label{eq: finite time state bound}
  \|x(t,j)\|\le M_T,
  \qquad (t,j)\in\operatorname{dom}q,\quad t\le T.
\end{equation}

Along flows, the membrane-potential equations imply
$\sum_{\ell,i,m}\dot\xi_{\ell,i,m}=\|x\|_1$,
because $\operatorname{ReLU}(v)+\operatorname{ReLU}(-v)=|v|$.
Each jump decreases $\sum_{\ell,i,m}\xi_{\ell,i,m}$ by exactly
$\Delta$.
Integrating over the flow intervals and accounting for the $j$
jumps gives
\begin{align}
  j\Delta
  &=\sum_{\ell,i,m}
    \bigl(\xi_{\ell,i,m}(0,0)-\xi_{\ell,i,m}(t,j)\bigr)
    +\int_0^t\|x(\sigma)\|_1\,d\sigma \notag\\
  &\le 2nN\Delta+\sqrt{nN}\,M_T T,
    \qquad (t,j)\in\operatorname{dom}q,\quad t\le T.
  \label{eq:finite-time-firing-bound}
\end{align}
Here the integral denotes the sum of the integrals over the
flow intervals.
Since $T>0$ was arbitrary, only finitely many jumps can occur
on any bounded time interval, excluding Zeno behavior.

It remains to prove completeness.
In the proof of Proposition~\ref{prop: well-posedness}, it is observed that $\mathcal H$ in \eqref{eq: hybrid MAS} satisfies
the hybrid basic conditions.
Let $T_{ C}(p)$ denote the tangent cone to $ C$ at $p$.
The definitions of $ C$, $ D$, and $F$ in \eqref{eq: hybrid MAS} give $F(p)\in T_{ C}(p)$ for every $p\in C\setminus D$.
Since $ D$ is closed,
$ C\setminus D$ is relatively open in $ C$.
Thus each $p\in C\setminus D$ has a neighborhood
$U$ such that $ C\cap U\subseteq C\setminus D$.
Therefore, we obtain
\begin{equation*}
  F(z)\cap T_{ C}(z)\ne\varnothing,
  \qquad z\in C\cap U.
\end{equation*}
Hence the viability condition in \cite[Proposition~6.10]{goebel2012hybrid} holds.
Since $G( D)\subset C\cup D$,
that proposition implies that $q$ is either complete or escapes to infinity in finite time along its last flow interval.
The latter is excluded by \eqref{eq: finite time state bound}, $\xi\in[0,\Delta]^{2nN}$, and $\tau(t,j)=t$.
Therefore, $q$ is complete. \hfill $\square$
\end{pf}

\subsection{Convergence to blended dynamics}

Let $\widetilde U\in\mathbb R^{N\times(N-1)}$ be any matrix so that $U:=\left[\frac{\mathbf 1_N}{\sqrt N}\ \ \widetilde U\right]$ is orthogonal.
Since $\mathcal{L}\mathbf 1_N=0$ and $\mathbf 1_N^\top \mathcal{L}=0$,
$U^\top \mathcal{L} U=\operatorname{diag}(0,\hat{\mathcal{L}})$, with $\hat{\mathcal{L}}:=\widetilde U^\top \mathcal{L}\widetilde U$ where the smallest eigenvalue of $\hat{\mathcal{L}}_s = \frac{\hat{\mathcal{L}} + \hat{\mathcal{L}}^\top}{2}$ is $\lambda_2$.
This provides, for $v\in\mathbb{R}^{N-1}$, $v^\top \hat{\mathcal{L}} v = v^\top \hat{\mathcal{L}}_s v \ge \lambda_2 \|v\|^2$.
Let
\begin{equation*}
    z_\mathsf{o} := \left[\frac{\mathbf{1}_N^\top}{N}\right]_{\otimes n} w,\quad z_\mathsf{x} := \widetilde U^\top_{\otimes n}w.
\end{equation*}
This yields $w = \mathbf{1}_N\otimes z_\mathsf{o} + \widetilde U_{\otimes n}z_\mathsf{x}$ and, with $\mathbf{1}_N^\top \mathcal{L}=0$,
\begin{equation}\label{eq: transformed dynamics}
\begin{aligned}
    \dot z_\mathsf{o} &= \left[\frac{\mathbf{1}_N^\top}{N}\right]_{\otimes n} f(t,x)\\
    \dot z_\mathsf{x}&= -k\hat{\mathcal{L}}_{\otimes n}z_\mathsf{x} + \widetilde U^\top_{\otimes n}f(t,x) + k \hat{\mathcal{L}}_{\otimes n}\widetilde U^\top_{\otimes n} (w-x),
\end{aligned}
\end{equation}
where the last term is bounded by $k\|{\mathcal{L}}\| c_w\alpha$, uniformly in
$(t,j)$, using Lemma~\ref{lem: jump invariance}.

The coordinate transformation indicates that $w$ consists of a synchronization component $z_\mathsf{o}$ and a disagreement component $z_\mathsf{x}$.
Since $x=\mathbf1_N\otimes z_\mathsf{o} + \widetilde U_{\otimes n}z_\mathsf{x} - (w-x)$,
and Lemma~\ref{lem: jump invariance} gives $\|w-x\|\le c_w \alpha$, $z_\mathsf{x}$ also measures
the disagreement of the physical state $x$ up to a $c_w \alpha$ reconstruction error.
If $z_\mathsf{x}=0$ and $w=x$, then $x=\mathbf{1}_N\otimes z_\mathsf{o}$, and the $z_\mathsf{o}$-dynamics
reduces exactly to the blended dynamics~\eqref{eq: Blended dynamics}, which is an emergent one.
We now impose a contraction condition on this candidate emergent dynamics.

\begin{asm}\label{asm: stability of blended dynamics}
The blended dynamics \eqref{eq: Blended dynamics} is contractive; that is, there exist a matrix $H_c\succ0$ and a constant $\lambda_c>0$ such that 
\begin{equation*}
H_c \partial_2 \bar f(t,s)  +  \left( \partial_2  {\bar f} \right) ^\top(t,s)H_c \preceq -2\lambda_c H_c,
\end{equation*}
for all $s\in\mathbb{R}^n$ and $t\ge0$~\cite{pavlov2005convergent}.
\end{asm}
This assumption ensures boundedness of the trajectory of the blended dynamics as follows.

\begin{lem}\label{lem: ultimate_bound}
Under Assumptions~\ref{asm: Regularity} and
\ref{asm: stability of blended dynamics}, let \(s(t)\) be the
solution of the blended dynamics~\eqref{eq: Blended dynamics}
with \(s(0)=s_0\). Define $\kappa(H_c):=\|H_c\|\|H_c^{-1}\|$.
Then, for every \(s_0\in\mathbb R^n\) and \(t\ge0\),
\begin{equation}\label{eq: blended uniform Euclidean bound}
    \|s(t)\| \le \sqrt{\kappa(H_c)}\|s_0\|e^{-\lambda_c t} + R_\infty
    \left(1-e^{-\lambda_c t}\right),
\end{equation}
where
    $R_\infty:=\sqrt{\kappa(H_c)}/\lambda_c \cdot m_0$.
\end{lem}

\begin{pf}
The proof is a straightforward extension of the scalar case established in \cite{kim2015robustness}.
By defining the Lyapunov candidate $V_s = \|s\|_{H_c}$ and applying the integral form of the mean-value theorem
\begin{equation}\label{eq: MVT}
\mu(t,a) - \mu(t,b) = \left( \int_0^1 \partial_2 \mu(t, b + \tau (a-b) ) \, d\tau \right) (a-b),
\end{equation}
the remainder of the analysis parallels the exact arguments in \cite{kim2015robustness} to yield the bound \eqref{eq: blended uniform Euclidean bound}.
\hfill $\square$
\end{pf}

The following theorem presents the main convergence analysis of the proposed spike-based communication framework, characterizing the ultimate boundedness of the synchronization error with respect to the blended dynamics trajectory.

\begin{thm}\label{thm: main}
Under Assumptions~\ref{asm: connected}--\ref{asm: stability of blended dynamics}, there exist constants $k^\star\ge0$,
$\beta_1>0$, and $\beta_2,\beta_3\ge0$, such that,
for every $k>k^\star$, $\alpha>0$, and $\Delta=\alpha/k$,
every maximal solution $q$ to $\mathcal H$ in \eqref{eq: hybrid MAS} with $q(0,0)\in C\cup D$ and $\tau(0,0)=0$ satisfies
\begin{equation}
\label{eq:ultimate-bound-main}
  \limsup_{\substack{(t,j)\in\operatorname{dom}q\\t\to\infty}}
  \|x_i(t,j)-s(t)\|
  \le\beta_1\alpha+\frac{\beta_2+\beta_3\alpha}{k-k^\star},
  \quad i\in\mathcal N,
\end{equation}
for every solution $s$ of the blended dynamics with arbitrary $s(0)\in\mathbb R^n$.
\end{thm}

\begin{pf}
Fix a maximal solution $q=(x,\xi,\tau)$ to $\mathcal H$ in
\eqref{eq: hybrid MAS} and an arbitrary solution $s$ of the
blended dynamics.
By Proposition~\ref{prop: complete-zeno}, the projection of
$\operatorname{dom}q$ onto the continuous-time axis is
$\mathbb R_{\ge0}$.
Let $e:=z_{\mathsf{o}}-s$ and $r:=w-x$.
Then
\begin{equation}
\label{eq:state-reconstruction-main}
  x=1_N\otimes(s+e)+\widetilde U_{\otimes n}z_{\mathsf{x}}-r,
  \qquad \|r\|\le c_w\alpha.
\end{equation}
Since $e$ and $z_{\mathsf{x}}$ are unchanged at jumps and $q$ is
Zeno-free, they define locally absolutely continuous functions
of $t\ge0$, denoted by the same symbols.

The error satisfies
\begin{align*}
  \dot e
  &=\bar f(t,s+e)-\bar f(t,s)\\
  &\quad+\left[\frac{1_N^\top}{N}\right]_{\otimes n}
    [f(t,x)-f(t,1_N\otimes(s+e))].
\end{align*}
Assumption~\ref{asm: stability of blended dynamics} and the
integral mean-value theorem~\eqref{eq: MVT} imply
\begin{equation*}
  e^\top H_c[\bar f(t,s+e)-\bar f(t,s)]
  \le-\lambda_c\|e\|_{H_c}^2.
\end{equation*}
For $e\ne0$, Cauchy--Schwarz in the $H_c$-metric,
$\|(1_N^\top/N)_{\otimes n}\|=1/\sqrt N$, and
Assumption~\ref{asm: Regularity} therefore give
\begin{align}
  D^+\|e\|_{H_c}
  &=\frac{e^\top H_c\dot e}{\|e\|_{H_c}} \notag\\
  &\le-\lambda_c\|e\|_{H_c} \notag\\
  &\quad+\frac{\sqrt{\|H_c\|}}{\sqrt N}
    \|f(t,x)-f(t,1_N\otimes(s+e))\| \notag\\
  &\le-\lambda_c\|e\|_{H_c}
    +a(\|z_{\mathsf{x}}\|+c_w\alpha),
  \label{eq:average-comparison-main}
\end{align}
where $a:=\sqrt{\|H_c\|}L_f/\sqrt N$ and the last step uses
\eqref{eq:state-reconstruction-main} and
$\|\widetilde U_{\otimes n}\|=1$.
The same inequality holds at $e=0$ in the upper Dini derivative sense.

Define the projected heterogeneity
\begin{equation*}
  m(t):=\|\widetilde U^\top_{\otimes n}
             f(t,1_N\otimes s(t))\|,
\end{equation*}
which is zero if the vector fields $f_i$ are identical.
For $z_{\mathsf{x}}\ne0$, using the disagreement dynamics in
\eqref{eq: transformed dynamics},
$z_{\mathsf{x}}^\top\widehat{\mathcal L}_{\otimes n}
z_{\mathsf{x}}\ge\lambda_2\|z_{\mathsf{x}}\|^2$,
\eqref{eq:state-reconstruction-main}, and
Assumption~\ref{asm: Regularity}, we obtain
\begin{align}
  D^+\|z_{\mathsf{x}}\|
  &\le-k\lambda_2\|z_{\mathsf{x}}\|
    +L_f(\sqrt N\|e\|+\|z_{\mathsf{x}}\|+\|r\|) \notag\\
  &\quad+m(t)+k\|\mathcal L\|\|r\| \notag\\
  &\le-d_k\|z_{\mathsf{x}}\|+b\|e\|_{H_c}
    +m(t)+\phi_{k,\alpha},
  \label{eq:disagreement-comparison-main}
\end{align}
where
\begin{gather*}
  d_k:=k\lambda_2-L_f,\qquad
  b:=\sqrt{N\|H_c^{-1}\|}\,L_f,\\
  \phi_{k,\alpha}:=(k\|\mathcal L\|+L_f)c_w\alpha.
\end{gather*}
The same inequality also holds at $z_\mathsf{x}=0$ in the upper Dini derivative sense.

Moreover, Assumption~\ref{asm: Regularity} and the bound on $s$
from Lemma~\ref{lem: ultimate_bound} imply
\begin{equation}
\label{eq:heterogeneity-bound-main}
  \sup_{t\ge0}m(t)<\infty,\qquad
  \overline m :=\limsup_{t\to\infty}m(t)\le\overline M,
\end{equation}
with $\overline M:=\sqrt N(L_fR_\infty+m_0)$.
Here, $\overline m$ does not depend on $s(0)$: for two blended solutions $s_1,s_2$,
Assumption~\ref{asm: stability of blended dynamics} gives $\|s_1(t)-s_2(t)\|\le\sqrt{\kappa(H_c)}\,e^{-\lambda_c t}\|s_1(0)-s_2(0)\|$~\cite{pavlov2005convergent},
while $|m(t;s_1)-m(t;s_2)|\le\sqrt{N}L_f\|s_1(t)-s_2(t)\|$ by Assumption~\ref{asm: Regularity}.

Set
\begin{equation}
\label{eq:sufficient-gain-main}
  k^\star:=\frac{L_f+ab/\lambda_c}{\lambda_2}
  =\frac{L_f}{\lambda_2}
    \left(1+\frac{L_f\sqrt{\kappa(H_c)}}{\lambda_c}\right),
\end{equation}
and let $k>k^\star$.
Then $d_k>ab/\lambda_c\ge0$ and
$\chi_k:=\lambda_c d_k-ab
=\lambda_c\lambda_2(k-k^\star)>0$, so
\begin{equation*}
  A_k:=\begin{bmatrix}
    -\lambda_c&a\\
    b&-d_k
  \end{bmatrix}
\end{equation*}
is Metzler and, having negative trace and positive determinant,
Hurwitz.
Consequently, $e^{A_kt}\ge0$ entrywise for $t\ge0$, and
\begin{equation*}
  \int_0^\infty e^{A_kt}\,dt=-A_k^{-1}
  =\frac{1}{\chi_k}
    \begin{bmatrix}d_k&a\\b&\lambda_c\end{bmatrix}\ge0.
\end{equation*}
Let
\begin{equation*}
  \eta:=\begin{bmatrix}
    \|e\|_{H_c}\\
    \|z_{\mathsf{x}}\|
  \end{bmatrix},
  \qquad
  u:=\begin{bmatrix}
    ac_w\alpha\\
    m+\phi_{k,\alpha}
  \end{bmatrix}.
\end{equation*}
By \eqref{eq:average-comparison-main} and
\eqref{eq:disagreement-comparison-main},
$\dot\eta=A_k\eta+u-\rho$ almost everywhere for some locally
integrable $\rho\ge0$.
Since $e^{A_kt}\ge0$, the variation-of-constants formula gives,
componentwise, for all $t\ge T\ge0$,
\begin{equation}
\label{eq:positive-comparison-main}
\begin{aligned}
  \eta(t)
  &\le e^{A_k (t-T)}\eta(T)
    +\int_T^t e^{A_k(t-\sigma)}u(\sigma)\,d\sigma\\
  &\le e^{A_k (t-T)}\eta(T)-A_k^{-1}\bar u,
\end{aligned}
\end{equation}
where $\bar u:=\sup_{\sigma\ge T}u(\sigma)$ componentwise, which is finite by \eqref{eq:heterogeneity-bound-main}; the last step uses $u\ge0$ and $e^{A_k\sigma}\ge0$.
Letting $t\to\infty$ and then $T\to\infty$, with $A_k$ Hurwitz
and $-A_k^{-1}\ge0$, yields
\begin{equation}\label{eq:comparison-ultimate-main}
  \limsup_{t\to\infty}\eta(t)
  \le-A_k^{-1}
    \begin{bmatrix}
      ac_w\alpha\\
      \overline m+\phi_{k,\alpha}
    \end{bmatrix}
  =:\begin{bmatrix}
    \overline E\\
    \overline Z
  \end{bmatrix}
\end{equation}
componentwise.
Evaluating $-A_k^{-1}$ and using
$\lambda_2k^\star=L_f+ab/\lambda_c$ gives
\begin{equation*}
  \overline Z
  =\frac{\overline m+
         c_w\alpha(k\|\mathcal L\|+\lambda_2k^\star)}
        {\lambda_2(k-k^\star)},
  \qquad
  \overline E
  =\frac{a}{\lambda_c}(\overline Z+c_w\alpha).
\end{equation*}

Since $\widetilde U\widetilde U^\top=I_N-1_N1_N^\top/N$,
every block row of $\widetilde U_{\otimes n}$ has norm
$c_U:=\sqrt{1-1/N}$.
Thus \eqref{eq:state-reconstruction-main} gives, for every
$(t,j)\in\operatorname{dom}q$,
\begin{equation*}
  \|x_i(t,j)-s(t)\|
  \le c_H\|e(t)\|_{H_c}
      +c_U\|z_{\mathsf{x}}(t)\|+c_w\alpha,
\end{equation*}
where $c_H:=\sqrt{\|H_c^{-1}\|}$.
Therefore, \eqref{eq:comparison-ultimate-main} yields
\begin{equation*}
  \limsup_{\substack{(t,j)\in\operatorname{dom}q\\t\to\infty}}
  \|x_i(t,j)-s(t)\|
  \le c_w\left(1+\frac{c_Ha}{\lambda_c}\right)\alpha
      +c_z\overline Z,
\end{equation*}
where $c_z:=c_U+c_Ha/\lambda_c$.
Writing $k=(k-k^\star)+k^\star$ in $\overline Z$ proves
\eqref{eq:ultimate-bound-main} with
\begin{align}
  \beta_1
  &:=c_w\left(
    1+\frac{c_Ha}{\lambda_c}
     +\frac{c_z\|\mathcal L\|}{\lambda_2}
  \right), \notag\\
  \beta_2
  &:=\frac{c_z\overline m}{\lambda_2},
  \qquad
  \beta_3:=c_wc_zk^\star
    \left(1+\frac{\|\mathcal L\|}{\lambda_2}\right).
  \label{eq:beta-main}
\end{align}
All these constants are independent of $k$, $\alpha$, and
the initial conditions. \hfill $\square$
\end{pf}

\subsection{Certified accuracy--communication trade-off}

Theorem~\ref{thm: main} shows that the certified ultimate synchronization bound can be made arbitrarily small by increasing the coupling gain $k$ and decreasing the spike amplitude $\alpha$.
However, since the threshold is given by $\Delta = \alpha/k$, such a naive parameter selection drives $\Delta$ close to zero, thereby leading to denser firing, which increases the communication burden and may pose implementation difficulties.

To quantify this trade-off, we rewrite the certified bound in Theorem~\ref{thm: main} in terms of the NM-AMP parameters $(\alpha,\Delta)$. Since $k=\alpha/\Delta$, for $\alpha,\Delta>0$ satisfying the sufficient gain condition $\alpha>k^\star\Delta$, the theorem yields the ultimate synchronization-error bound
\begin{equation}
\gamma(\alpha,\Delta) := \beta_1\alpha + \frac{\beta_2\Delta}{\alpha-k^\star\Delta} + \frac{\beta_3\alpha\Delta}{\alpha-k^\star\Delta}.
\end{equation}
We use this certified bound to derive parameter choices under two representative design constraints.

\textbf{Case 1 (Bandwidth-constrained design):} 
When communication bandwidth limits the minimum threshold potential ($\Delta \ge \Delta_\mathrm{min} > 0$), we can minimize the upper bound of synchronization error by solving:
\begin{equation}
    \begin{aligned}
        & \underset{\alpha, \Delta}{\text{minimize}}
        & & \gamma(\alpha, \Delta) \\
        & \text{subject to}
        & & \Delta_\mathrm{min} \le \Delta, \qquad \alpha > k^{\star}\Delta.
    \end{aligned}
\end{equation}
Accounting for the stability boundary, the exact analytical solution is given by
\begin{equation}
    \Delta^* = \Delta_\mathrm{min}, \quad \alpha^* = k^\star \Delta_\mathrm{min} + \sqrt{\frac{\Delta_\mathrm{min}(\beta_2 + \beta_3 k^\star\Delta_\mathrm{min})}{\beta_1}}.
\end{equation}
Here, $k^\star\Delta_{\min}$ identifies the lower bound associated with the sufficient gain condition, whereas the square-root term gives the optimal positive margin above that boundary.

\textbf{Case 2 (Performance-constrained design):}
For applications strictly requiring the synchronization error bound to be $\gamma(\alpha, \Delta) \le \Gamma$, we maximize the communication threshold $\Delta$ by solving
\begin{equation}
    \begin{aligned}
        & \underset{\alpha, \Delta}{\text{maximize}}
        & & \Delta \\
        & \text{subject to}
        & & \gamma(\alpha,\Delta)\le\Gamma, \qquad \alpha > {k^{\star}}\Delta.
    \end{aligned}
\end{equation}
The optimal closed-form parameters are explicitly given by
\begin{equation}
\begin{aligned}
\alpha^* &= \frac{\Gamma(\beta_2 + k^\star \Gamma)}{\beta_1(\beta_2+ k^\star\Gamma) + \sqrt{\beta_1 (\beta_2 + k^\star\Gamma)(\beta_1\beta_2 + \beta_3\Gamma)}} \\
\Delta^* &= \frac{\beta_1(\alpha^*)^2}{\beta_2 + k^\star\Gamma}.
\end{aligned}
\end{equation}

In Case 2, the explicit solution of $\Delta^*$ reveals how network heterogeneity dictates the communication cost. 
We first note that in practical scenarios requiring high-precision synchronization, the target accuracy $\Gamma$ is chosen to be sufficiently small.
In this regime, applying a first-order Taylor approximation yields $\alpha^* \approx \frac{1}{2\beta_1}\Gamma$, and consequently, $\Delta^* \approx \frac{1}{4\beta_1(\beta_2 + k^\star\Gamma)} \Gamma^2$.
Therefore, the optimal communication threshold $\Delta^*$ under the certified bound scales as follows:
\begin{itemize}
    \item If the heterogeneity-dependent coefficient $\beta_2$ is sufficiently large compared to the target accuracy such that $\beta_2 \gg k^\star\Gamma$, the denominator is dominated by $\beta_2$, yielding $\Delta^* \propto \Gamma^2$. In this regime, reducing the synchronization error by half imposes a strictly quadratic penalty, requiring a fourfold increase in the communication frequency;
    \item Conversely, if the network is nearly homogeneous such that $\beta_2 \ll k^\star\Gamma$ for a given target accuracy, the scaling law transitions to $\Delta^* \propto \Gamma$. In this case, halving the error linearly requires only a twofold increase in the communication frequency.
\end{itemize}

Beyond the two representative cases discussed above, the simple algebraic structure of the ultimate bound $\gamma(\alpha, \Delta)$ provides a highly tractable framework.
The simple algebraic structure of $\gamma(\alpha,\Delta)$ also facilitates other application-specific design problems, admitting analytical solutions in simple cases and numerical solutions in general.

\section{An illustrative example}

We illustrate the proposed spike-based communication
architecture using the distributed median solver
in \cite{lee2020fully}.
Consider a distributed median solver consisting of a network of $N= 5$ heterogeneous agents with
\begin{equation*}
f_i(x_i) = \operatorname{sgn}(c_i - x_i), \quad i \in \mathcal{N} = \{1, 2, 3, 4, 5\},
\end{equation*}
where $(c_1,c_2,c_3,c_4,c_5) = (2,2,11,1,2)$.
The agents are interconnected through the directed ring shown in Fig.~\ref{fig: Midsol topology}, which is strongly connected and balanced.

The corresponding blended dynamics is
\begin{equation*}
  \dot s = \frac{1}{5}\sum_{i=1}^{5}\operatorname{sgn}(c_i-s) \in -\frac{1}{5}\partial V(s),
  \quad
  V(s):=\sum_{i=1}^{5}|s-c_i|
\end{equation*}
where $\partial$ is the subdifferential operator.
Since $N=5$ is odd, $V$ has the unique minimizer: $\operatorname{median}\{c_1,\ldots,c_5\}=2.$

\begin{figure}[t]
    \centering
    \begin{tikzpicture}
    
        \foreach \i in {1,...,5} {
            \node[draw, circle, inner sep=1pt, color=agentcolor\i, thick] 
              (agent\i) at ({72*\i+18}:3) {agent \i};
            \node (spike\i) at ($(agent\i)+({72*\i-90}:1.2)$) {};
          }
        \foreach \i in {1,...,5}{
        \pgfmathtruncatemacro{\j}{mod(\i,5)+1}
            \draw[->] (agent\i) -- (agent\j) node[pos = 0.33, fill=white, draw, rectangle, inner sep=2pt, sloped] {\tiny NM-AMP};
        }

        \foreach \i in {1,...,5}{
        \pgfmathsetmacro{\sign}{ifthenelse(mod(\i,2)==0,1,-1)}
            \draw[-] ($(spike\i) - (0.7cc,0)$) -- ($(spike\i) + (0.7cc,0)$);
            \draw[-] ($(spike\i) - (0.2cc,0)$) -- ($(spike\i) - (0.2cc, \sign*0.3cc)$) node[pos=1.0, draw, circle, fill=black, inner sep=0.01cc]{};
            \draw[-] ($(spike\i) - (0.4cc,0)$) -- ($(spike\i) - (0.4cc,0.3cc)$) node[pos=1.0, draw, circle, fill=black, inner sep=0.01cc]{};
            \draw[-] ($(spike\i) - (0.5cc,0)$) -- ($(spike\i) - (0.5cc,\sign*0.3cc)$) node[pos=1.0, draw, circle, fill=black, inner sep=0.01cc]{};
            \draw[-] ($(spike\i) - (0.3cc,0)$) -- ($(spike\i) - (0.3cc, -0.3cc)$) node[pos=1.0, draw, circle, fill=black, inner sep=0.01cc]{};
            \draw[-] ($(spike\i) + (0.2cc,0)$) -- ($(spike\i) + (0.2cc,\sign*0.3cc)$) node[pos=1.0, draw, circle, fill=black, inner sep=0.01cc]{};
            \draw[-] ($(spike\i) + (0.4cc,0)$) -- ($(spike\i) + (0.4cc,0.3cc)$) node[pos=1.0, draw, circle, fill=black, inner sep=0.01cc]{};
            \draw[-] ($(spike\i) + (0.3cc,0)$) -- ($(spike\i) + (0.3cc, -0.3cc)$) node[pos=1.0, draw, circle, fill=black, inner sep=0.01cc]{};
            \draw[-] ($(spike\i)$) -- ($(spike\i) + (0, 0.3cc)$) node[pos=1.0, draw, circle, fill=black, inner sep=0.01cc]{};
        }

        \def\givenVal{{2, 2, 11, 1, 2}}

        \foreach \i in {1, 2, 3}{
        \pgfmathsetmacro{\val}{\givenVal[\i-1]}
            \draw[-{Triangle}, thick] ($(agent\i)-(1.5,0)$) -- (agent\i) node[pos =0.4, above] {\val};
        }
        
        \foreach \i in {4,5}{
        \pgfmathsetmacro{\val}{\givenVal[\i-1]}
            \draw[-{Triangle}, thick] ($(agent\i)+(1.5,0)$) -- (agent\i) node[pos =0.4, above] {\val};
        }

        \node (consensus) {\large $\approx$ 2};

        \foreach \i in {1, ..., 5}{
            \draw[-{Triangle}, thick, color = agentcolor\i] (agent\i) -- (consensus);
        }
\end{tikzpicture}
    \caption{Network topology of distributed median solver enabled by neuro-spike communication}
    \label{fig: Midsol topology}
\end{figure}

\begin{figure}[t]
\centering
\includegraphics[width=1\linewidth]{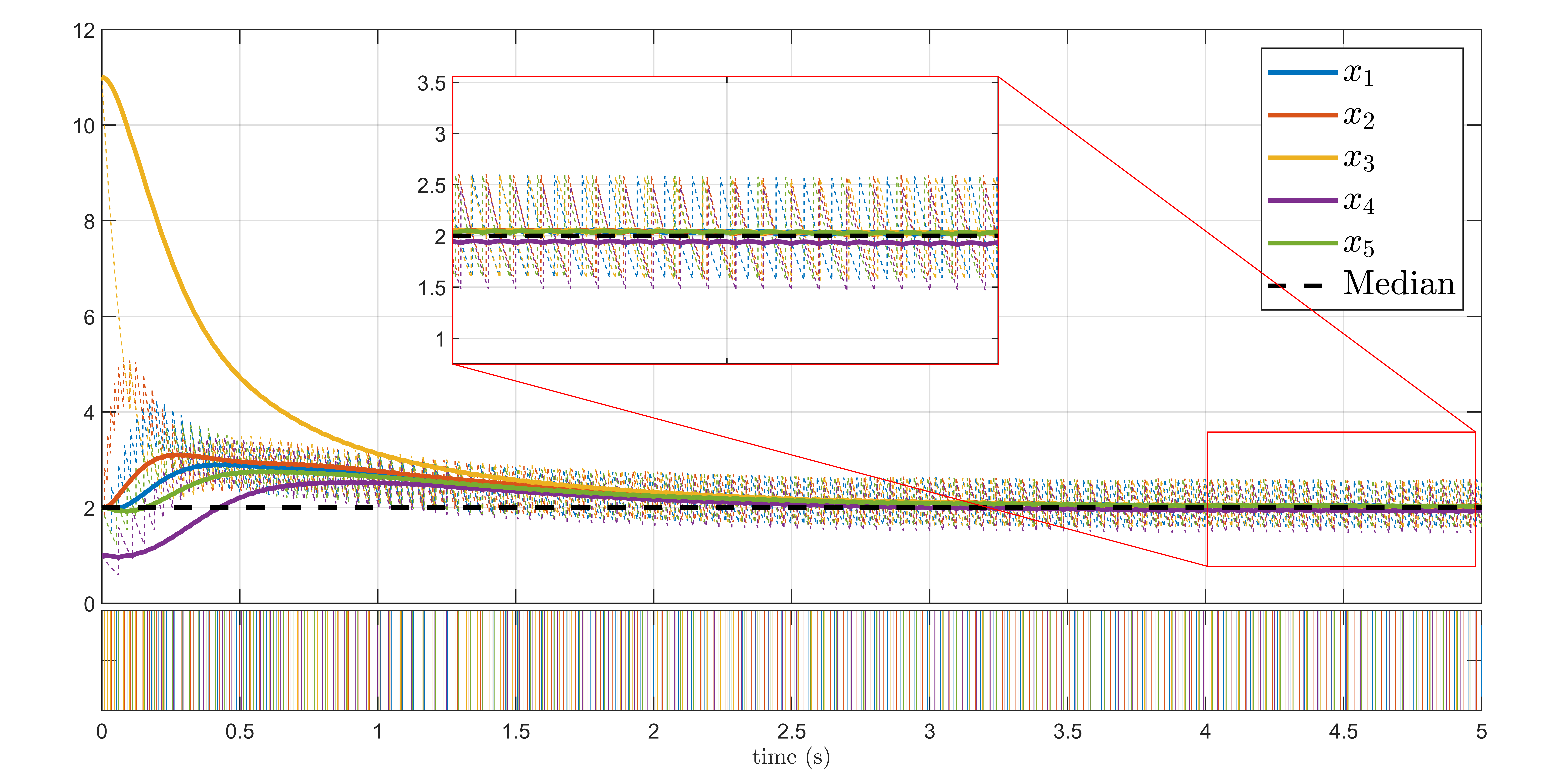}
\caption{Raw and filtered state trajectories of the five agents, together with their communication instants ($\alpha=1,\Delta=0.1$).}
\label{fig: med. sol.}
\end{figure}

The upper panel of Fig.~\ref{fig: med. sol.} shows that the trajectories approach a neighborhood of the median $2$, despite the outlying reference $c_3=11$. The dashed curves show the raw state trajectories, including the spike-induced jumps, while the thick solid curves show the corresponding outputs of a first-order low-pass filter with the time constant $0.3$.
The filter is used only for visualization: its output does not enter the closed-loop dynamics or the firing logic.
Note that the filter attenuates the spike-induced fluctuations, making the role of the spike amplitude $\alpha$, as highlighted by Theorem~\ref{thm: main}, less apparent in the filtered trajectories.
The inset surrounded by red square magnifies the steady-state trajectories.
The lower panel displays the firing times of the five agents, illustrating the intermittent and asynchronous nature of the communication.
Counting both polarities, a total of $564$ spikes are emitted over the $5\,\mathrm{s}$ simulation interval.
This corresponds to an average firing rate of
\begin{equation*}
  \frac{564}{5\times5} =22.56 \quad\text{spikes/s per agent}.
\end{equation*}

This distributed median solver example is chosen to examine whether the proposed neuro-spike implementation can reproduce a meaningful collective behavior in a setting beyond the assumptions of Theorem~\ref{thm: main}.
For example, the vector fields are discontinuous, and the blended dynamics does not satisfy the smooth contraction condition in Assumption~\ref{asm: stability of blended dynamics}.
Thus, Theorem~\ref{thm: main} does not guarantee the observed behavior.
The simulation shows that the trajectories nevertheless approach a neighborhood of the median, suggesting that the proposed implementation may retain this collective behavior beyond the theorem’s scope.
Establishing this observation rigorously is left for future work.

\section{Conclusion}

This paper developed a neuro-spike communication framework for emergent behavior in heterogeneous multi-agent systems, in which inter-agent information is conveyed only through intermittent and asynchronous signed spikes.
By formulating the closed-loop network as a hybrid system and introducing a coordinate invariant under spike-induced jumps, we removed the jump contributions from the convergence estimate and reduced the main convergence analysis to the flow dynamics.
Under the stated connectivity, regularity, and contraction assumptions, we established that every maximal solution is Zeno-free and complete and that the agents practically synchronize while collectively following a trajectory of the blended dynamics.
The resulting explicit ultimate bound characterizes a certified accuracy--threshold trade-off and yields closed-form parameter choices under minimum-threshold and prescribed-accuracy constraints.
A distributed median solver, although outside Assumption~\ref{asm: Regularity} and \ref{asm: stability of blended dynamics}, numerically illustrates the potential applicability of the proposed architecture beyond the analyzed setting.
Furthermore, the proposed framework is applicable to broader distributed tasks, such as network size estimation, optimization, and observer design, as discussed in \cite{lee2021design}.

\begin{ack}                               
This work was supported by
the National Research Foundation of Korea (NRF) grant
funded by the Korea government (MSIT) (No. RS-2026-25504174).
H. Jang thanks Jinwook Heo for helpful discussions that improved the proof of Theorem~\ref{thm: main}.
\end{ack}

\section*{Declaration of generative AI and AI-assisted technologies in the manuscript preparation process}
During the preparation of this work, the authors used Chat GPT and Claude in order to improve the language and readability of the manuscript.
After using this tool/service, the authors reviewed and edited the content as needed and take full responsibility for the content of the published article.

\bibliographystyle{plain}        
\bibliography{ref}               

\end{document}